\documentclass[prl,twocolumn,showpacs,floatfix,superscriptaddress]{revtex4}
\usepackage{graphicx,amsfonts,amssymb,amsmath, hyperref}
\usepackage[applemac]{inputenc}

\newif\ifhyper
\hypertrue
\ifhyper
\hypersetup{
  citecolor = {green},
  colorlinks = {true}, 
  urlcolor = {blue} 
}
\fi

\newlength{\ldag}
\begin{document}

\title{Crumpled-to-tubule  transition in anisotropic polymerized  membranes:  beyond $\epsilon$-expansion}

\author{K. Essafi} 
\email{essafi@lptmc.jussieu.fr}
\affiliation{LPTMC,
CNRS UMR 7600, UPMC, 4 Place Jussieu, 75252 Paris Cedex 05, France}

\author{J.-P.  Kownacki}
\email{kownacki@u-cergy.fr}
\affiliation{LPTM, CNRS UMR 8089-Universit\'e de Cergy-Pontoise,   2 avenue Adolphe Chauvin, 95302 Cergy-Pontoise  Cedex, France}

\author{D. Mouhanna} 
\email{mouhanna@lptmc.jussieu.fr}
\affiliation{LPTMC,
CNRS UMR 7600, UPMC, 4 Place Jussieu, 75252 Paris Cedex 05, France}


\begin{abstract}
Anisotropic  $D$-dimensional polymerized phantom membranes are investigated  within a nonperturbative renormalization group (NPRG) framework.  One  focuses  on the transition between  a high-temperature, crumpled,  phase and   a   low-temperature,  tubular,  phase  where the  membrane is flat along one direction and crumpled along the other ones.   While the upper critical dimension -- $D_{\hbox{\scriptsize uc}}=5/2$ -- is   close to  $D=2$  the weak-coupling perturbative approach is   qualitatively and quantitatively {\it wrong}. We show that our  approach is free of the problems encountered within the perturbative framework  and provides physically meaningful critical quantities.  

\end{abstract}

\pacs{87.16.D-, 11.10.Hi, 11.15.Tk}

\maketitle

Polymerized phantom membranes  display  a remarkable behavior as the temperature is varied \cite{bowick01,proceedings89}. While crumpled at high temperature  due to fluctuations, they exhibit  a low-temperature  flat phase with  long-range orientational order  \cite{nelson87,david88,paczuski88}.   The existence of an ordered phase in $D=2$  despite Mermin-Wagner theorem originates in  the existence of an anharmonic   coupling   between   bending and  elastic degrees of freedom or, equivalently,  between   out-of-plane capillary  and in-plane phonon, modes \cite{nelson87}.   This  coupling  induces a   phonon-mediated   long-range effective interaction between the out-of-plane fluctuations that  stabilizes  a low-temperature  flat phase    for two -- and even less -- dimensional membranes \cite{aronovitz88,guitter89,aronovitz89}.   Within a  RG  picture  the  coupling  between the different modes is responsible  for   a stiffening  of the bending rigidity  constant at low-momentum  -- $\kappa(q)\sim q^{-\eta}$ with $\eta>0$  -- called anomalous elasticity, that  suppresses   the destructive   fluctuations  in the transverse direction to the membrane \cite{nelson87,aronovitz88,aronovitz89}.  Note that,   although strongly reduced, the out-of-plane  fluctuations are not completely suppressed and  the membrane  still display  {\it ripples}  \cite{meyer07,fasolino07,guinea08}.

This phenomenon of  long-range order induced by  coupling between bending and elastic degrees of freedom  has   provided  the  theoretical grounds  for  the existence and stability  of  polymerized membranes  and,  incidentally,   of  the recently discovered   graphene \cite{novoselov04},  the first example of truly  two-dimensional  membrane.  It has   also  underlined  the importance of the nature of the membrane internal structure  regarding   the kind of  order  displayed at long distance.   As a remarkable illustration of this fact is, in marked contrast with the case of polymerized  membranes,  the absence  of long-range orientational order  for membranes  deprived of  fixed connectivity, and thus of  elasticity modulus,  whose prototypical example are  liquid membranes.  Past   studies have considered  a wide variety  of  alterations of  the conventional crystalline order of polymerized membranes: presence of  topological defects, like dislocations or disclinations,   vacancies, local variations of connectivity,  random impurities, $\dots$   giving  rise  to novel  and rich critical  behaviors including new  universality classes (see \cite{bowick01}  and \cite{radzihovsky04} for reviews).

\begin{figure}[htbp]
\begin{center}{\includegraphics[width=0.48\textwidth]{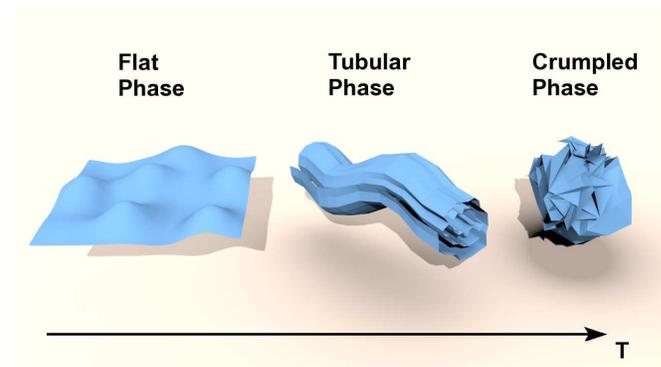}}
\end{center}
\caption{Flat, tubular and crumpled phases of anisotropic membranes as functions of  the temperature.}
\label{tubule}
\end{figure}

A  fruitful  modification   of   the   translationally/rotationally invariant internal structure of  the membrane  is the inclusion  of an  in-plane  {\it anisotropy}.   In \cite{radzihovsky95,radzihovsky98}  it has been shown that  membranes with intrinsic one-directional anisotropy  display, when the temperature is lowered,  a {\it tubular} phase,   flat  in one direction and  crumpled   in the other ones. This phase lies in a   temperature range intermediate between a high-temperature regime, corresponding to a fully  crumpled phase and a low-temperature regime,  where the membrane adopts a flat conformation  in all directions, see Fig.\ref{tubule}.

While the existence and stability of  tubular phases have been numerically confirmed \cite{bowick97,koibuchi10},   the crumpled-to-tubule transition remains numerically and experimentally unexplored.  The  main reason for this situation  is the lack of  accurate predictions  for  the critical quantities. This is  {\it a priori}  a rather surprising situation.  Indeed, a striking  consequence  of the presence  of  anisotropy is that  the upper critical dimension is lowered from $D_{\hbox{\scriptsize uc}}=4$, for  isotropic membranes, to $D_{\hbox{\scriptsize uc}}=5/2$. From the proximity of $D_{\hbox{\scriptsize uc}}$ with   $D=2$  one could expect the  perturbative approach performed in  the parameter $\epsilon=5/2-D$ to be in  a good position  to evaluate  the critical quantities  in $D=2$.  However,  as shown explicitly by Radzihovsky and Toner,  due to  the fractional nature  of the upper critical dimension,   the  $\epsilon$-expansion is, on the contrary,  "{\it extremely}  unreliable" and even "{\it qualitatively}  wrong" \cite{radzihovsky98}.  This is clear  in view of  the exponent $\eta$ describing the  correlation between the tangents of the membrane, which   is found to be {\it negative}. A genuine  negative value for this exponent  would  correspond  to  a    downward renormalization -- a decreasing -- of  the bending rigidity constant, in contradiction with what is expected from physical grounds \cite{radzihovsky98}. Moreover  alternative perturbative  methods  used to investigate  the low-energy physics of membranes, {\it e.g.}  the self consistent screening approximation (SCSA),  that have successfully worked  for  isotropic membranes \cite{ledoussal92,gazit09,zakharchenko10},  appear  here  hard to implement   given the complexity of the field theoretical formulation of the model.

In this article we investigate the crumpling-to-tubule phase transition of anisotropic membranes   by means of a NPRG approach.  This method \cite{wetterich93c}   has been recently used to investigate  the  crumpling-to-flat phase transition and the flat phase of  $D$-dimensional isotropic membranes  embedded in a $d$-dimensional space \cite{kownacki09} (see also \cite{braghin10}).     One of its  important features   is that it  does not rely on the proximity of an  upper or lower critical dimension.     This is the reason why it has allowed to confidently  investigate the physics of membranes in the  whole $(D,d)$ plane \cite{kownacki09}. Such a method is  appropriate    in the present context where  the $\epsilon$-expansion   displays a pathological behavior.

The NPRG approach relies on the use of a running effective action  \cite{wetterich93c}  (see \cite{bagnuls01,berges02, delamotte03,pawlowski07,rosten10}  for reviews),  $\Gamma_k[\bf r]$,   a functional of    $\bf r=r({\bf x})$    a $d$-dimensional   {\sl  external}  vector  that  describes the membrane  in the embedding space, ${\bf x}$   being   a  set of   {\sl  internal}  $D$-dimensional  coordinates  labeling a point  of the membrane.   The index  $k$  stands for  a running scale that separates the high-momentum,  with $q>k$,  from  the low-momentum ones, with $q<k$ and $\Gamma_k[\bf r]$  represents   a coarse grained free energy where only fluctuations with momenta $q\ge k$  have been integrated out.  The  running  of $k$ towards the  value  $k=0$ thus corresponds to  gradually integrate more and more  low-momentum fluctuations. The $k$-dependence, RG flow,  of $\Gamma_k$ is  provided  by an exact evolution equation \cite{wetterich93c}:
\begin{equation}
{\partial \Gamma_k\over \partial t}={1\over 2} \hbox{Tr} \left\{(\Gamma_k^{(2)}+R_k)^{-1}
 {\partial R_k\over \partial t}\right\}
\label{renorm}
\end{equation}
where $t=\ln \displaystyle {k / \Lambda}$, $\Lambda$ being some microscopic, lattice, scale. The trace in (\ref{renorm}) involves  a $D$-dimensional momentum  integral as well as a summation over vectorial  indices. The function $R_k(q)$  realizes  the split between low- and high-momentum degrees of freedom. Several forms  of $R_k(q)$ will be considered in the following.  Finally $\Gamma_k^{(2)}$ is the second functional derivative  of  $\Gamma_k[\bf r]$ with respect to $\bf r$.

 The effective action   $\Gamma_k[\bf r]$  relevant to study polymerized membranes must be invariant under the group of translations, which  implies  that it  only depends on powers of the tangent vectors $\partial_{\mu}\bf{r}$.    For anisotropic membranes rotational invariance {\it within} the membrane  is explicitly broken between  one  direction -- named $y$ -- and the remaining  $D-1$  dimensions which are  kept  isotropic.     The effective  action thus writes:
\begin{equation}
\begin{array}{ll}
\Gamma_k[{\bf r}] = \displaystyle \int d^{D-1}x~dy &\Bigg\{\displaystyle {Z_y \over 2}(\partial_y^2{\bf r})^2 + t_{\perp}(\partial_{\mu}^{\perp}{\bf r})^2 \\
&\displaystyle +\,  {u_y\over 2}(\partial_y {\bf r}. \partial_y {\bf r}-\zeta_y^2)^2 \Bigg\}
\label{effectiveaction}
 \end{array}
\end{equation}
where   $Z_y$, $u_y$, $t_{\perp}$ and $\zeta_y$ are running  coupling constants:  $Z_y$  is  a bending rigidity, $u_y$  an elastic constant and   $t_{\perp}$  a tension term  in the $\perp$-directions (indexed by $\mu$ that runs from 1 to $D-1$). This last term, that plays the role of a temperature in the $\perp$-directions,  is kept  non-vanishing since these directions  are not critical.  Finally  $\zeta_y$,  or rather the temperature along $y$,  $t_y\equiv u_y\, \zeta_y^2$,  parametrizes the approach to criticality in the $y$-direction.  Note that, up to change of coupling constants, this action is  that used in \cite{radzihovsky95,radzihovsky98}.  Let  us consider the mean-field treatment of  this model for $u_y>0$.  When $\zeta_y=0$, the minimum of $\Gamma_k$  is given by a configuration where $\partial_y {\bf r}$ vanishes which characterizes a crumpled, disordered,  phase.  When  $\zeta_y >0$,  this minimum is given by a configuration ${\bf r}({\bf x}) = \zeta_y\,  y\,  {\bf e_{\hbox{\tiny D}}}$ where  ${\bf e_{\hbox{\tiny D}}}$ is a unit vector that spans   the  one-dimensional submanifold along which   long-range order   occurs.   Thus,  when the  temperature  $t_y$   is lowered,  the system undergoes a phase transition between a crumpled phase at high temperature, with $\zeta_y=0$,  and a tubular phase  at a low, intermediate,  temperature,  with $\zeta_y\ne 0$.  When the temperature is further lowered,  one recovers   an isotropic  flat phase  along  all  the  $D$-directions, see Fig.(\ref{tubule}). A specificity of the model is that at  the crumpling-to-tubule transition  power counting leads to  $q_{\perp}\propto q_y^2$,  an  anisotropic scaling  which characterizes a {\it Lifshitz-type} behavior \cite{hornreich75}.  According to this scaling behavior the most relevant terms are those entering in  Eq.(\ref{effectiveaction}).  Note that since we  treat  nonperturbatively this action we are not supposed to base our computation  on power-counting arguments. We do this  nevertheless since  {\it i)}  the proximity of $D=2$  with  the upper critical dimension  $D_{\hbox{\scriptsize uc}}=5/2$ leads to guess that the terms present in Eq.(\ref{effectiveaction}) play the major role  {\it  ii)}  we wish  to compare our approach  with the previous perturbative ones \cite{radzihovsky95,radzihovsky98}. The  assumption   {\it i)}  must, of course, be  checked, which can be partly done by  evaluating the sensitivity of  the results with respect to modification of the cut-off  function $R_k(q)$. 

Let us now define the critical quantities.  First, due to the anisotropy between the $y$-direction and the remaining $D-1$  $\perp$-directions one 
has the  scale transformations:  $x_{\perp}  = k\,  x'_{\perp}$ and $y  =  k^z\,  y' \  $ that define the anisotropic scaling exponent $z$. Under a RG transformation the field $\bf r$ acquires an anomalous dimension $\eta$ such that $Z_y\sim k^{-\eta\,  z}$, while   $t_{\perp}\sim k^{-\eta_{\perp}}$ so  the field $\bf r$ generically scales as:  ${\bf r}= k^{{1\over 2}{(D-1-3 z+\eta z)}}\, {\bf r'}$. One deduces  the following relation between $z$, $\eta$ and $\eta_{\perp}$: $z=(2-\eta_{\perp})/( 4-\eta)$.
Finally one defines the  exponents $\nu_y$ and $\nu_{\perp}$  from the correlation lengths near criticality:  $\xi_y\propto t_y^{-\nu_y}$ and $\xi_{\perp}\propto t_y^{-\nu_{\perp}}$ with $\nu_y=z\, \nu_{\perp}$ \cite{radzihovsky98}.  The flow equations for the couplings  $Z_y$, $u_y$, $t_{\perp}$ and $\zeta_y$ are obtained  using their definitions in terms of functional derivatives of the effective action and applying  the RG Eq.(\ref{renorm}). One defines the dimensionless quantities: $\overline\zeta_y^2={k_y^{3-2D} Z_y^{3-D\over 2} t_{\perp}^{D-1\over 2}} \zeta_y^2$ and $\overline u_y={k_y^{2D-5} Z_y^{D-5\over 2} t_{\perp}^{1-D\over 2}} u_y$ with $k_y=k^z$. In terms of these quantities the RG equations write, with $t=\displaystyle \ln {k_y / \Lambda}$:
\begin{eqnarray}
\partial_t  t_{\perp} &=&  0 \nonumber  \\
 \nonumber  \\
\partial_t \overline \zeta_y^2 &=& -\left(2D-3-\eta {(D-3)/ 2}\right)\overline\zeta_y^2 \nonumber  \\ 
 \nonumber \\
& &-\,  (d-1)\,  \overline l_{{3-D\over 2},0}^{2,D}
 - 3\  \overline l_{0,{3-D\over 2}}^{2,D}    \nonumber    \\
 \label{eqrg} \\
\partial_t \overline u_y &=& -\left(5-2D-\eta {(5-D)/2}\right)\overline u_y   + \nonumber \\
\nonumber \\
&& (D-3)\, \overline u_y^2\, \left[(d-1)\,  \overline l_{{5-D\over 2},0}^{4,D}+ 9\,   \overline l_{0,{5-D\over 2}}^{4,D}\right]   \nonumber
\end{eqnarray}
while the anomalous dimension {\it function} $\eta$  writes:
\begin{eqnarray}
\eta  &=& \frac{1}{3}(D-3)\,\overline u_y^2\,\overline \zeta_y^2\Big[-108\,   \overline l_{0,{5-D\over 2}}^{2,D} - 12(d-1)\,  \overline l_{{5-D\over 2},0}^{2,D}  \nonumber \\
 \nonumber \\
&&-540 (D-5)\, \overline u_y\,\overline \zeta_y^2\,  \overline l_{0,{7-D\over 2}}^{4,D}   \nonumber \\
&& +  (D-5)(D-7) \left( -288\,  \overline u_y^2\, \overline \zeta_y^2 \,  \overline l_{0,{9-D\over 2}}^{6,D} +9 \,  \overline M_{0,{9-D\over 2}}^{6,D}  \right.  \nonumber \\
 \nonumber \\
&& \left. +(d-1)\,  \overline M_{{9-D\over 2},0}^{6,D}  -36\, \overline u_y\, \overline \zeta_y^2\,  \overline N_{0,{9-D\over 2}}^{6,D} \right)\Big] \ .
 \label{eta}  
\end{eqnarray}
In Eqs.(\ref{eqrg}) and (\ref{eta})  $\overline  l_{a, b}^{\alpha, D}$ , $\overline  N_{a, b}^{\alpha, D} $ and $\overline M_{a, b}^{\alpha, D}$ are  the de-dimensionalized  "threshold functions"  (see \cite{berges02,delamotte03})   $l_{a, b}^{\alpha, D}$,  $N_{a, b}^{\alpha, D} $ and $M_{a, b}^{\alpha, D}$  that are  given by :
\begin{equation}
  T_{a, b}^{\alpha, D} = K_D\  \displaystyle  \widehat{\partial \over \partial t} \int dq_y\,  q_y^{\alpha}\,  {F[q_y]\over \left[P(q_y)\right]^a \left[P(q_y)+  m_y^2\,  q_y^2\right]^b}
\label{threshold}
\end{equation}
where $K_D= \left({\pi/ 2}\right)^{{(D-1)/ 2}}\Gamma\left[{(3-D)/ 2}\right]$, $P(q_y)=Z_y\, q_y^{4}+R_{k}(q_y)$,   $m_y^2=4  u_y  \zeta_y^2$ and where $\widehat{{\partial}}/{\partial t}$ only acts on
$R_{k}$. In Eq.(\ref{threshold})  the function  $F[q_y]$ is given by $1$, $dP/dq_y^2$ and $(dP/dq_y^2)^2$ for $l$, $N$ and $M$ respectively.  These functions  control  the relative  weight of the different   modes:   the -- single -- phonon mode of mass $m_y$  and the $d-1$ capillary modes of zero mass,  along   the RG flow.   Note, that these functions  only depend on $q_y$.  Indeed  $\partial_{\mu}^{\perp}$ enters   only quadratically in the action (\ref{effectiveaction}) so that  one can exactly perform the integration on the $D-1$ $\perp$-degrees of freedom in the $\beta$ functions. 

Let us now discuss  our results.  First  we find  no renormalization of $t_{\perp}$, i.e. $\eta_{\perp}=0$, for any value of $D$,    in agreement with the all-order perturbative result  \cite{radzihovsky98}.  Due to the one-loop structure of the  RG equation (\ref{renorm})   we recover the one-loop $\beta$ function for $ \overline u_y$ and $t_y= \overline u_y  \overline \zeta_y^2$ found in \cite{radzihovsky98} using  our equations (\ref{eqrg}) expanded around $D=5/2-\epsilon$:
\begin{eqnarray}
\partial_t \overline \zeta_y^2 &=&   \displaystyle  - 2\,   \overline\zeta_y^2 +  {2\over 3} (d+2)\,  K_{5/2} \nonumber     \\
\partial_t \overline u_y &=&   \displaystyle - 2\,  \epsilon\,  \overline u_y + (d+8) \  K_{5/2}  \, \overline u_y^2\nonumber 
\end{eqnarray}
with, at this order,   a vanishing anomalous dimension $\eta^*$.  Thanks again to  the  one-loop structure of (\ref{renorm}), one  gets   the large-$d$ results from the set of equations (\ref{eqrg})-(\ref{eta}), without almost any more computation.   In this limit, assuming that  $ \overline u_y\sim O(1/d)$ and $ \overline \zeta_y^2\sim O(d)$ one finds, in agreement  with this assumption,  and using, e.g., the cut-off function
$R_k(q_y)=Z_y (k^4 - q_y^4) \theta(k^2 - q_y^2)$,  a stable fixed point with coordinates: 
\begin{equation}
\begin{array}{ll}
{ \overline\zeta_{y}^2}^*&=\displaystyle{4d\over 3}\,  {(3-D)\over (2D-3)} K_D \\
\\
 \overline u_y^*&=\displaystyle {5\over 4 d}{(5-2D)\over (3-D)(5-D)K_D} \ . 
\end{array}
\end{equation}
This fixed point exists for all values of $D$ between $5/2$ and the lower critical dimension $D=3/2$ (see \cite{radzihovsky98}). The corresponding critical exponents are given by: $\nu=1/(2D -3) +O(1/d)$ and $\eta=O(1/d)$.  
 
\begin{figure}[htbp]
\begin{center}{\includegraphics[width=0.35\textwidth]{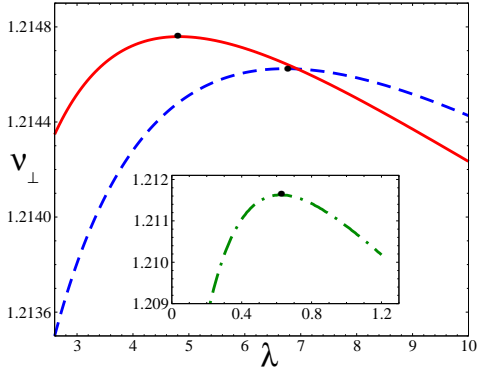}}
\end{center}
\caption{The exponent $\nu_{\perp}$ as function of the parameter $\lambda$. In solid line $ \widetilde R_k^1(q_y)$, in  dashed line   $ \widetilde R_k^2(q_y)$ and  in dot-dashed  line  $ \widetilde R_k^3(q_y)$. The dot on each curve corresponds to a stationary value of the exponent.}
\label{fignu}
\end{figure}

\begin{figure}[htbp]
\begin{center}{\includegraphics[width=0.35\textwidth]{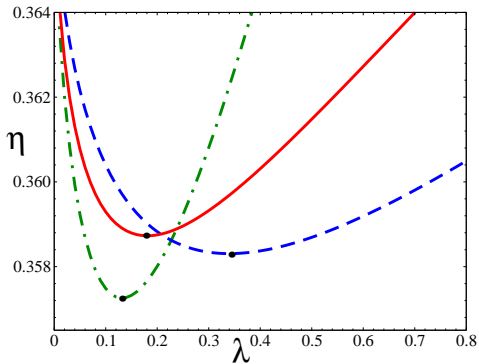}}
\end{center}
\caption{The exponent $\eta$ as function of the parameter $\lambda$. The conventions are the same as in Fig.(\ref{fignu}).}
\label{figeta}
\end{figure}

Going to  finite values of $d$ one finds a stable fixed point  for any value of  $d$  in contrast to the isotropic case where it occurs only for  $d$ greater that some critical value \cite{paczuski88} $d_{cr}(D)$   determined in \cite{kownacki09}. One  now specializes to the $d=3$ case.  One finds, for {\it any}  dimension $D<5/2$,  a nontrivial anomalous dimension $\eta^*$ from Eq.(\ref{eta}) at the fixed point.   This exponent  is found {\it positive}  in any dimension  between  $D=5/2$  and  $D=3/2$, in agreement with what is expected.   The critical exponents $\eta^*$ and $\nu_{\perp}$  in $D=2$  are displayed in Figs.(\ref{fignu})-(\ref{figeta}). They are represented  as functions  of a  real number   $\lambda$ that parametrizes  three families of cut-off functions   among those  that we have used: $R_k^i(q)=\lambda \widetilde R_k^i(q)$, $i=1,2,3$ with  $\widetilde R_k^1(q_y)=Z_y k^4 \hbox{exp}(-q_y^4/k^4)$, $ \widetilde R_k^2(q_y)=Z_y /  (\hbox{exp}(q_y^4/k^4)-1)$  and  $ \widetilde R_k^3(q_y)=Z_y (k^4 - q_y^4) \theta(k^2-q_y^2)$.  Variations of $\lambda$ allow  to investigate the cut-off dependence of the critical quantities and to optimize each  cut-off function inside its family \cite{canet03a}, {\it i.e.} to (try to) find stationary values  of  the critical quantities.  For each exponent one succeeds to find  a -- single  -- stationary point  for closed values  of $\lambda$, see Figs.(\ref{fignu})-(\ref{figeta}).  For  $\nu_{\perp}$ one can extract from these considerations the value $\nu_{\perp}=1.213(8)$    in agreement with that found in \cite{radzihovsky98}:  $\nu_{\perp}\simeq 1.227$. For the anomalous dimension one finds  $\eta^*\simeq 0.358(4)$ which  differs  considerably from the value found in  \cite{radzihovsky98}: $\eta\simeq -0.0015$.  Finally one deduces from these results and the scaling relations:   $z= 0.5490(6)$ and $\nu_y=0.665(5)$ to be compared to $\nu_y\simeq 0.614$. Note that  variations of the critical quantities  appear to be  very smooth with both variations  {\it i)} of the parameter $\lambda$ inside a family of cut-off function and  {\it ii)}  of  the family itself, see Figs.(\ref{fignu})-(\ref{figeta}).   This  large   insensitivity with respect  to  the cut-off functions constitutes a strong indication of trustability of our results, in agreement with  the proximity of the upper critical dimension.

 We  have  shown that the NPRG   allows to  overcome the main difficulty  that plague the  perturbative, weak coupling,   expansion for  anisotropic membranes.  Our  predictions  can be easily  tested  through  numerical  investigations.  This would  validate the NPRG approach as an efficient alternative  to the perturbative one, in particular in the perspective of probing the  properties of  membrane-like  systems  with various orders  and geometries.

\acknowledgments{We thank C. Bervillier,  B. Delamotte, N. Dupuis  for helpful discussions.}

\end{document}